\documentclass[12pt]{article}
\usepackage{graphicx}
\def\hybrid{\topmargin 0pt      \oddsidemargin 0pt
        \headheight 0pt \headsep 0pt
       \voffset-1cm
        \textwidth 6.25in       
       \textheight 9.5in       
        \marginparwidth 0.0in
        \parskip 5pt plus 1pt   \jot = 1.5ex}
\catcode`\@=11
\def\marginnote#1{}

\newcount\hour
\newcount\minute
\newtoks\amorpm
\hour=\time\divide\hour by60
\minute=\time{\multiply\hour by60 \global\advance\minute by-\hour}
\edef\standardtime{{\ifnum\hour<12 \global\amorpm={am}%
        \else\global\amorpm={pm}\advance\hour by-12 \fi
        \ifnum\hour=0 \hour=12 \fi
        \number\hour:\ifnum\minute<10 0\fi\number\minute\the\amorpm}}
\edef\militarytime{\number\hour:\ifnum\minute<10 0\fi\number\minute}

\def\draftlabel#1{{\@bsphack\if@filesw {\let\thepage\relax
   \xdef\@gtempa{\write\@auxout{\string
      \newlabel{#1}{{\@currentlabel}{\thepage}}}}}\@gtempa
   \if@nobreak \ifvmode\nobreak\fi\fi\fi\@esphack}
        \gdef\@eqnlabel{#1}}
\def\@eqnlabel{}
\def\@vacuum{}
\def\draftmarginnote#1{\marginpar{\raggedright\scriptsize\tt#1}}

\def\draftlabel#1{{\@bsphack\if@filesw {\let\thepage\relax
   \xdef\@gtempa{\write\@auxout{\string
      \newlabel{#1}{{\@currentlabel}{\thepage}}}}}\@gtempa
   \if@nobreak \ifvmode\nobreak\fi\fi\fi\@esphack}
        \gdef\@eqnlabel{#1}}
\def\@eqnlabel{}
\def\@vacuum{}
\def\draftmarginnote#1{\marginpar{\raggedright\scriptsize\tt#1}}

\def\draft{\oddsidemargin -.5truein
        \def\@oddfoot{\sl preliminary draft \hfil
        \rm\thepage\hfil\sl\today\quad\militarytime}
        \let\@evenfoot\@oddfoot \overfullrule 3pt
        \let\label=\draftlabel
        \let\marginnote=\draftmarginnote
   \def\@eqnnum{(\theequation)\rlap{\kern\marginparsep\tt\@eqnlabel}%
\global\let\@eqnlabel\@vacuum}  }

\def\numberbysection{\@addtoreset{equation}{section}
        \def\theequation{\thesection.\arabic{equation}}}

\def\underline#1{\relax\ifmmode\@@underline#1\else
        $\@@underline{\hbox{#1}}$\relax\fi}

\def\titlepage{\@restonecolfalse\if@twocolumn\@restonecoltrue\onecolumn
     \else \newpage \fi \thispagestyle{empty}\c@page\z@
        \def\thefootnote{\fnsymbol{footnote}} }

\def\endtitlepage{\if@restonecol\twocolumn \else  \fi
        \def\thefootnote{\arabic{footnote}}
        \setcounter{footnote}{0}}  
\relax

\hybrid

\newfont{\Bbb}{msbm10 scaled 1\@ptsize00}
\newfont{\Bbbb}{msbm7 scaled 1\@ptsize00}

\newcommand{\DDD}{\raise-1pt\hbox{$\mbox{\Bbbb D}$}}

\newcommand{\UUU}{\raise-1pt\hbox{$\mbox{\Bbbb U}$}}

\newcommand{\ZZ}{\mbox{\Bbb Z}}
\newcommand{\z}{\raise-1pt\hbox{$\mbox{\Bbbb Z}$}}

\def\beq{\begin{equation}}
\def\eeq{\end{equation}}
\def\p{\partial}

\def\l{\lambda}

\newcommand{\bpmatrix}{\left ( \begin{array}{c}  }
\newcommand{\epmatrix}{\end{array} \right )  }

\def\square{\hfill
{\vrule height6pt width6pt depth1pt} \break \vspace{.01cm}}

\newtheorem{theorem}{Theorem}[section]
\newtheorem{lemma}{Lemma}[section]
\newtheorem{lemma-definition}{Lemma-Definition}[section]
\newtheorem{corollary}{Corollary}[section]

\newtheorem{remark}{Remark}[section]

\newtheorem{definition}{Definition}[section]
\newtheorem{proposition}{Proposition}[section]

\usepackage{color}

\begin{document}

\begin{titlepage}

\title{Formal solution to the KP hierarchy}

\author{S.M.~Natanzon\thanks{National Research University Higher School of Economics, 20 Myasnitskaya Ulitsa,
Moscow 101000, Russia and ITEP 25 B.Cheremushkinskaya, Moscow 117218, Russia , e-mail: natanzons@mail.ru}
\and A.V.~Zabrodin
\thanks{Institute of Biochemical Physics,
4 Kosygina st., Moscow 119334, Russia; ITEP, 25
B.Cheremushkinskaya, Moscow 117218, Russia and
Laboratory of Mathematical Physics,
National Research University Higher School of Economics,
20 Myasnitskaya Ulitsa,
Moscow 101000, Russia, e-mail: zabrodin@itep.ru}}

\date{September 2015}
\maketitle

\vspace{-7cm} \centerline{ \hfill ITEP-TH-16/15}\vspace{7cm}

\begin{abstract}
We find all formal solutions to the
$\hbar$-dependent KP hierarchy. They are characterized by certain Cauchy-like data.
The solutions are found in the form 
of formal series for the tau-function of the hierarchy and for its logarithm
(the $F$-function).
An explicit combinatorial description of the
coefficients of the series is provided.
\end{abstract}


\end{titlepage}

\tableofcontents


\vspace{1ex}

\section{Introduction}

\subsection{Motivation}

The partial differential equation (PDE)
\beq\label{intro1}
\frac{3}{4}\, u_{yy}=
\frac{\p}{\p x}\left ( u_t -\frac{3}{2}\, u u_x -\frac{1}{4}\, u_{xxx}
\right )
\eeq
was derived by Kadomtsev and Petviashvili (KP) in 1970 for description of
non-linear waves in two-dimensional media with small dispersion.
Later it was recognized that this equation
is integrable and it can be naturally embedded
into an infinite system of compatible PDE's. This system is now
called the KP integrable hierarchy. The KP hierarchy can be most naturally
formulated in terms of bilinear equations for the tau-function
$\tau = \tau (t_1, t_2, t_3, t_4, \ldots )$ of the infinite
set of ``times'' ${\bf t} =\{t_1, t_2, t_3, \ldots \}$, the first
three of which are identified with $x,y,t$ as $t_1=x$, $t_2=y$,
$t_3=t$. The variable $u$ in equation (\ref{intro1}) is expressed
through the tau-function as $u=2\p_x^2F$, where $F=\log \tau$.

The KP hierarchy has a lot of exact solutions of very different
nature: non-linear waves and excitations in
dispersive media (quasi-periodic and soliton solutions \cite{Kr77,Nov80}),
generating functions
for topological invariants in algebraic geometry \cite{Wit91,Lan02},
partition functions for models
of random matrices \cite{Mor94}, etc.

The tau-function admits the well-known
expansion in Schur functions depending on the times $t_i$ \cite{Sato}. The structure
of this expansion is studied in detail and well understood \cite{DJKM83,JM83,EH10}.
However,
in many cases it is the $F$-function which is of prime interest
rather than the tau-function
itself. At the same time, series expansions for the $F$-function are missing
in the literature. This is the problem that we address in the present paper.
It should be noted that we deal with
not necessarily convergent series. The corresponding solutions are
called formal.

One can introduce an auxiliary formal parameter $\hbar$ by re-scaling
$t_k\to t_k/\hbar$, then the KP equation acquires the form
\beq\label{intro2}
\frac{3}{4}\, u_{yy}=
\frac{\p}{\p x}\left ( u_t -\frac{3}{2}\, u u_x -\frac{\hbar^2}{4}\, u_{xxx}
\right )
\eeq
The corresponding re-scaling in the whole hierarchy is called the
$\hbar$-formulation of the KP hierarchy (or the
$\hbar$-dependent KP hierarchy \cite{TakTak}). The coefficients
of its formal solutions are formal series in $\hbar$.
The $F$-function is defined as $F=\hbar ^2 \log \tau$.

The parameter
$\hbar$ has a meaning of the Planck's constant in the ``auxiliary space'',
i.e., the space where the Lax operator of the KP hierarchy acts.
The limit $\hbar \to 0$ is the quasiclassical limit in the auxiliary space.
In the ``physical space'' of dynamical variables this limit is realized
as the dispersionless limit, where higher space-time derivatives
of dynamical variables are
set to be negligibly small. For example, equation (\ref{intro2})
at $\hbar =0$ becomes the Khokhlov-Zabolotskaya equation
(the dispersionless KP equation).

The $\hbar \to 0$ limit of the KP hierarchy is called
the dispersionless KP (dKP) hierarchy. It is a particular case
of general Whitham hierarchy of PDE's
\cite{KriW1,KriW}. The dispersionless limit is of great
interest on its own. (For applications to interface
dynamics and problems of complex analysis
see \cite{MWWZ,MWZ99,ZJPA,Z,NZ} and \cite{WZ,Ztmf}
respectively.) A class of solutions to the dKP hierarchy
can be obtained as the $\hbar \to 0$ limits of solutions to the
$\hbar$-dependent KP hierarchy. However, such limits do not always
exist in the class of smooth functions.

In the present paper we construct all formal solutions to the $\hbar$-dependent
KP hierarchy in the form of an infinite formal series with the coefficients
defined by an explicit recurrence procedure.
The solution is determined by an infinite set of arbitrary functions of one
variable $f_0(x), \, f_1(x), \, f_2 (x),\, \ldots$ which are supposed to be
formal series or
differentiable infinitely many times. These functions are initial data for the
solution. In particular, $F(x;{\bf 0})=f_0(x)$. However,
the functions $f_i$ with $i\geq 1$ coincide with the standard Cauchy data
$\p_{t_i}F(x, {\bf t} )\Bigm |_{{\bf t}=0}$ only at $\hbar =0$.
At $\hbar \neq 0$ these functions differ from the first order derivatives
by some terms which are of higher order in $\hbar$. We call the set
$\{f_0(x), \, f_1(x), \, f_2 (x),\, \ldots \}$ {\it the Cauchy-like data}.

The fact that
solutions to the KP hierarchy can be restored from the first order
derivatives of the $F$-function at ${\bf t}=0$
was pointed out earlier \cite{DN,N01,N03a}. However, no explicit
solution of the Cauchy problem for the KP hierarchy is available even
in the sense of formal series.
In this paper
we give explicit formulas that allow one to restore the formal
solution from the arbitrary set
of the Cauchy-like data $\{f_0(x), \, f_1(x), \, f_2 (x),\, \ldots \}$.


\subsection{Preliminaries}


Here we list
the necessary notations and facts related to Young diagrams and Schur functions.

\paragraph{Young diagrams.}
Following \cite{Macdonald}, we will
denote the Young diagrams as $\lambda$, $\mu$, etc.
Let $\lambda =[\lambda_1,\lambda_2,\ldots,\lambda_\ell ]$ be
the Young diagram with $\ell =\ell (\lambda )$
rows of non-zero lengths $\lambda_1 \geq \lambda_2
\geq \ldots \geq \lambda_\ell >0$.
We identify $\lambda$ with the partition of the number
$|\lambda |:=\lambda_1 +\ldots +
\lambda_{\ell}$ into the $\ell$ non-zero parts $\lambda_i$.
Another convenient notation is $\lambda =(1^{m_1}2^{m_2}\ldots
r^{m_r}\ldots )$, which means that exactly $m_i$ parts of the
partition $\lambda$ have length $i$: $m_i=\mbox{card}\, \{j: \lambda_j=i\}$.
In particular, the Young diagram with one row (respectively, one column) of length $k$ is
denoted as $(k)$ (respectively, $(1^k)$).
Put
$$
\rho(\lambda )=\lambda_1\ldots
\lambda_{\ell (\lambda )}\,, \quad
\sigma (\lambda )=\prod_{k\geq 1}m_k!\,, \quad
z_{\lambda}=\sigma (\lambda )\rho (\lambda ).
$$

One can introduce some orderings on the set of diagrams with fixed
$|\lambda |$ \cite{Macdonald}. One of them is the reverse
lexicographical order: $\lambda$ precedes $\mu$ if the first
non-vanishing difference $\lambda_i -\mu_i$ is positive.
In this ordering $(n)$ comes first and $(1^n)$ comes last.
It is a linear ordering.
Another ordering is the natural partial ordering which is defined as follows:
\beq\label{ordering}
\lambda \geq \mu \,\,\,\, \Longleftrightarrow \,\,\,\,
\lambda_1 +\ldots +\lambda_i \geq
\mu_1+\ldots +\mu _i \qquad \mbox{for all $i\geq 1$}.
\eeq
As soon as $|\lambda |\geq 6$ it is not a total ordering.

\paragraph{Schur functions.}
The general reference is \cite{Macdonald}.
Let ${\bf t}=\{t_1, t_2, t_3, \ldots \}$ be an infinite set
of variables.
The Schur polynomials $s_{\lambda}({\bf t})$
labeled by Young diagrams $\lambda$ can be defined
by the determinant formula
\beq\label{T1}
s_{\lambda}({\bf t})=\det_{i,j=1, \ldots , \ell (\lambda )}
h_{\lambda_i -i +j}({\bf t}),
\eeq
where the polynomials $h_j({\bf t})$ are defined with the help of
the generating series
$$
\exp \Bigl ( \sum_{k\geq 1}t_k z^k\Bigr )=\sum_{k\geq 0} h_k({\bf t})z^k,
$$
or, explicitly,
$\displaystyle{
h_k ({\bf t})=\sum_{k_1+2k_2+\ldots =k}
\frac{t_1^{k_1}}{k_1!}\, \frac{t_2^{k_2}}{k_2!}\ldots
=\sum_{l=1}^{k}\frac{1}{l!}\! \sum_{{k_1,
\ldots , k_l \geq 1}\atop{k_1+\ldots +k_l=k}}\!\! t_{k_1}\ldots t_{k_l}\,.}
$
The first few polynomials are
$h_1({\bf t})=t_1$,
$h_2({\bf t})=\frac{1}{2}\, t_1^2 +t_2$,
$h_3({\bf t})=\frac{1}{6}t_1^3 +t_1t_2 +t_3$,
$h_4({\bf t})=\frac{1}{24}t_1^4 +\frac{1}{2}\, t_2^2 +
\frac{1}{2}\, t_1^2 t_2 +t_1t_3 +t_4$.
It is convenient to put $h_0({\bf t})=1$, $h_{k}({\bf t})=0$ for $k<0$
and $s_{\emptyset}({\bf t})=1$.
The functions $h_k$ are elementary Schur polynomials in the sense that
for one-row diagrams
$s_{(j)}({\bf t})=h_j ({\bf t})$. Equation (\ref{T1}) is known as the Jacobi-Trudi
identity.

We will need the Cauchy-Littlewood
identity
\beq\label{id}
\sum_{\lambda}s_{\lambda}({\bf t})s_{\lambda}({\bf t}')=
\exp \Bigl ( \sum_{k\geq 1}kt_k t'_k\Bigr ),
\eeq
where the sum is over all Young diagrams including the empty one.
The both sides can be regarded as formal series in the variables ${\bf t}, {\bf t}'$.
Writing it in the form
$$
\sum_{\lambda}s_{\lambda}({\bf y})s_{\lambda}(\tilde \p )=
\exp \Bigl ( \sum_{k\geq 1}y_k \p_{t_k}\Bigr ),
$$
where $\tilde \p =\{\p_{t_1} , \frac{1}{2}\p_{t_2}, \frac{1}{3}\p_{t_3},
\ldots \, \}$
and applying to $s_{\mu}({\bf t})$, we get the relation
\beq\label{id1}
\left. \phantom{\int}
s_{\lambda}(\tilde \p )s_{\mu}({\bf t})\right |_{{\bf t}=0}=
\delta_{\lambda \mu}
\eeq
which reflects the orthonormality of the Schur functions.

The Schur functions are usually regarded as symmetric functions
of the variables $x_i$ defined by $kt_k =\sum_i x_i^k$.
In terms of the variables $x_i$ ($i=1,\ldots , N$, $N\geq \ell (\lambda )$),
\beq\label{id2}
s_{\lambda}({\bf t})=s_{\lambda}(\{x_i\}_N)=
\frac{\det _{1\leq i,j\leq N}
\Bigl ( x_{i}^{N+\lambda_j -j}\Bigr )}{\det _{1\leq i,j\leq N}
\Bigl ( x_{i}^{N -j}\Bigr )}\,, \quad t_k =\frac{1}{k}\sum_i x_i^k\,.
\eeq
It can be proved \cite{Macdonald} that the Schur
polynomials form an orthonormal basis in the space of symmetric functions.

\paragraph{Dual bases in the space of polynomials.}
Let us consider the linear space ${\cal P}$ of polynomials in the
variables ${\bf t}=\{t_1, t_2, t_3, \ldots \}$.
The basis vectors are naturally labeled by
Young diagrams
with basis
polynomials $v_{\lambda}({\bf t})$ being
quasi-homogeneous in the sense that
$v_{\lambda}(a t_1, a^2 t_2, a^3 t_3,
\ldots )\! =\! a^{|\lambda |} v_{\lambda}(t_1, t_2, t_3, \ldots )$.
By definition, the polynomial labelled by the empty diagram $\emptyset$
is $v_{\emptyset}({\bf t})=1$.
One may choose such a basis in many different ways.
The simplest example is the usual monomial basis
$$
t_{\lambda}({\bf t}):=t_{\lambda_1}t_{\lambda_2}\ldots t_{\lambda_\ell}=
\prod_{i\geq 1}t_{i}^{m_i(\lambda )}
$$
Other examples are the basis
$h_{\lambda}({\bf t})=h_{\lambda_1}h_{\lambda_2}
\ldots h_{\lambda_{\ell}}$ and
the Schur polynomial basis $s_{\lambda}({\bf t})$.
Having in mind the interpretation of polynomials of $t_k$ as symmetric functions
of the variables $x_i$ discussed above,
we will call polynomials
from the space ${\cal P}$ symmetric functions (although they are {\it not}
symmetric w.r.t. the variables $t_k$). Another standard basis in
${\cal P}$ is the basis of monomial symmetric functions
$m_{\lambda}=m_{\lambda}({\bf t})$ which are easily defined in terms of the variables
$x_1, x_2, \ldots , x_n$ provided
$n\geq \ell (\lambda )$ (it is implied
that $\lambda_j=0$ if $j>\ell (\lambda )$):
$$
m_{\lambda}(x_1, x_2, \ldots , x_n)
=\frac{1}{(n-\ell (\lambda ))!\,
\sigma (\lambda )}\sum_{P\in S_n} x_1^{\lambda_{P(1)}}
x_2^{\lambda_{P(2)}}\ldots x_n^{\lambda_{P(n)}}.
$$
The normalization factor is chosen in such a way
that the function
$m_{\lambda}({\bf t})$ does not depend on $n$
if $n\geq \ell (\lambda )$.
For example, $m_{(k)}({\bf t})=\sum_j x_j^k =kt_k$.
The first few functions $m_{\lambda}$ are:
\beq\label{dual15a}
\begin{array}{l}
m_{(1)}({\bf t})=t_1
\\  \\
m_{(2)}({\bf t})=2t_2, \quad m_{(1^2)}({\bf t})=\frac{1}{2}\, t_1^2 - t_2
\\  \\
m_{(3)}({\bf t})=3t_3, \quad
m_{(21)}({\bf t})=2t_2 t_1 -3t_3, \quad
m_{(1^3)}({\bf t})=\frac{1}{6}\, t_1^3 - t_2 t_1 +t_3
\end{array}
\eeq
It is also convenient to keep the standard notation $p_k$ for the
power sums $p_k=kt_k$ and the corresponding basis
$p_{\lambda}({\bf t})=\rho (\lambda )t_{\lambda}$.

One can introduce the scalar product in the space ${\cal P}$ as follows:
\beq\label{dual1}
\left <u_{\lambda}, \, v_{\mu}\right >=
u_{\lambda}(\tilde \p )\, v_{\mu }({\bf t})\Bigr |_{{\bf t}=0}.
\eeq
This definition is symmetric,
i.e., $\left <u_{\lambda}, \, v_{\mu}\right >=
\left <v_{\mu}, \, u_{\lambda}\right >$.
This scalar product coincides with the one
from \cite{Macdonald} defined axiomatically in the space of
symmetric functions. In particular, we have:
$\left <p_{\lambda}, p_{\mu}\right >=z_{\lambda}\delta_{\lambda \mu}$,
where ,
$\left <s_{\lambda}, s_{\mu}\right >=\delta_{\lambda \mu}$
and $\left <h_{\lambda}, m_{\mu}\right >=\delta_{\lambda \mu}$.
If $\left <u_{\lambda}, v_{\mu}\right >=\delta_{\lambda \mu}$
for all $\lambda$, $\mu$, we say that $u_{\lambda}$,
$v_{\lambda}$ are dual bases.

\begin{lemma}
Let $u_{\lambda}({\bf t})$,
$v_{\lambda}({\bf t})$ be any pair of dual bases
in ${\cal P}$, and $g({\bf t})$
be any function real-analytic around the point
${\bf t}=0$, then the Taylor series at ${\bf t}=0$
can be represented in the form
\beq\label{dual2}
g({\bf t})=\sum_{\lambda}u_{\lambda}(\tilde \p )g({\bf t}')
\Bigr |_{{\bf t}'=0}\,
v_{\lambda}({\bf t}).
\eeq
\end{lemma}
\noindent
The proof is obvious. \square

\section{The $\hbar$-KP hierarchy for the tau-function}

Hereafter we work with
the $\hbar$-formulation \cite{TakTak} of the KP hierarchy and call it
the $\hbar$-KP hierarchy.
Let $\tau =\tau (t_1, t_2, t_3, \ldots )=\tau ({\bf t})$ be the tau-function of
the hierarchy. It depends on the infinite set of time variables
${\bf t}=\{t_1, t_2, \ldots \}$. It also contains $\hbar$ as a parameter but we will not
write it explicitly.

\subsection{Hirota equations for the tau-function}

The $\hbar$-KP hierarchy can be represented in different equivalent forms.
We start with the Hirota bilinear equations for the tau-function.
They can be encoded in a functional relation.
Below we use the notation
$$
\tau ^{[z_1, \ldots , z_m ]}({\bf t})=\tau \left ( {\bf t}+
\hbar \sum_{i=1}^m [z_i^{-1}]\right ) =
e^{\hbar (D(z_1)+\ldots +D(z_m))}\, \tau
$$
where
$$
\begin{array}{c}
{\bf t} \pm \hbar [z^{-1}]:=\Bigl \{ t_1 \pm \hbar z^{-1},
t_2 \pm \frac{\hbar}{2}\, z^{-2},
t_3 \pm \frac{\hbar}{3}\, z^{-3}, \ldots ,\Bigr \}
\end{array}
$$
and the differential operator $D(z)$ is defined by
\beq\label{DD}
D(z)=\sum_{k\geq 1}\frac{z^{-k}}{k}\, \p_k
\eeq
Hereafter we abbreviate $\p_k =\p / \p t_k$.

The Hirota functional relation for the tau-function reads
\beq\label{Hir1}
(z_1 \! -\! z_2) \tau ^{[z_1, z_2]}\tau ^{[z_3]}+
(z_2 \! -\! z_3) \tau ^{[z_2, z_3]}\tau ^{[z_1]}+
(z_3 \! -\! z_1) \tau ^{[z_3, z_1]}\tau ^{[z_2]}=0
\eeq
It is to be fulfilled for all $z_1, z_2, z_3$.
By tau-function (of the $\hbar$-KP hierarchy) we mean any solution to this equation.
Differential equations of the hierarchy are obtained by expanding
this equation in powers of $z_1, z_2, z_3$. Another form of the functional
relation for the $\tau$-function is
\beq\label{Hir2}
\hbar \p_{1}\log \frac{\tau ^{[z_1]}}{\tau ^{[z_2]}}=
(z_2-z_1) \left (\frac{\tau ^{[z_1, z_2]}\tau}{\tau ^{[z_1]}\tau ^{[z_2]}}-1\right )
\eeq
which is sometimes called the differential Fay identity.
\begin{proposition}
Equations (\ref{Hir1}) and (\ref{Hir2}) are equivalent.
\end{proposition}
{\it Proof.} Eq. (\ref{Hir2}) follows from eq. (\ref{Hir1}) in the limit $z_3\to \infty$.
Eq. (\ref{Hir1}) can be obtained from (\ref{Hir2}) by summing the equations written for the
pairs $\{z_1, z_2\}$,  $\{z_2, z_3\}$,  $\{z_3, z_1\}$.
\square

\begin{proposition}\label{taudet}
The tau-function $\tau = \tau ({\bf t})$ of the $\hbar$-KP hierarchy satisfies
the equations
\beq\label{Hir3}
\prod_{1\leq i<j\leq m}\! (z_j-z_i) \cdot
\tau ^{[z_1, \ldots , z_m]}\tau ^{m-1}=\det_{1\leq j,k\leq m}
\Bigl ((z_j \! -\! \hbar \p_1 )^{k-1}\tau ^{[z_j]}\Bigr )
\eeq
for any $m\geq 2$ and any $z_1, \ldots , z_m$.
\end{proposition}
{\it Proof.} At $m=2$ equation (\ref{Hir3}) reads
$$
(z_2-z_1)\tau ^{[z_1, z_2]} \tau =\left |
\begin{array}{ll} \tau ^{[z_1]} & (z_1\! - \! \hbar \p_1 )\tau ^{[z_1]}
\\ & \\
\tau ^{[z_2]} & (z_2\! - \! \hbar \p_1 )\tau ^{[z_2]} \end{array}
\right |
$$
which is (\ref{Hir2}).
The rest of the proof is induction in $m$. See Appendix A for details.
\square

\noindent
Equation (\ref{Hir3}) was first suggested in \cite{ALTZ14}.

\vspace{2ex}


\subsection{Formal solution for the tau-function}

Let us introduce the differential operators
\beq\label{dhbar}
\p^{\hbar}_{k}= \frac{k}{\hbar} \, h_k (\hbar \tilde \p ) =
\p_k + \hbar \sum_{l=1}^{k-1}\frac{k\p_l \p_{k-l}}{2l(k-l)}\, +
O(\hbar^2),
\eeq
where $h_k ({\bf t})$ are the elementary Schur polynomials.
These operators are ``$\hbar$-deformations''
of the partial derivatives in the sense that $\p^{\hbar =0}_k =\p_k$.
We call them $\hbar$-deformed partial derivatives.
Their properties are studied below in the next section.

In the KP theory the first variable, $t_1$, is distinguished.
Having this in mind,
we will treat the tau-function depending on $t_i$ as a result
of the evolution of $\tau (x; {\bf 0})$ according to the KP flows:
$\tau (x;{\bf 0}) \to
\tau (x;{\bf t})=f(x) \hat \tau (x+t_1, t_2, t_3 , \ldots )$. Our aim is to
represent it as a formal series in the $t_i$'s given $\tau (x;{\bf 0})$ and
some Cauchy-like data to be specified below.

With the help of the operators $\p_i^{\hbar}$ we now define
modified Cauchy data which are necessary for constructing the
formal solution. We call them the {\it Cauchy-like data}.
\begin{definition}\label{def-Cauchy-like}
The Cauchy-like data of a solution $\tau (x;{\bf t})$
to the $\hbar$-KP hierarchy is the
set of functions
$\tau (x;{\bf 0})$,
$\p_k^{\hbar}\tau (x;{\bf t})\Bigm |_{{\bf t}=0}$, $k\geq 1$.
\end{definition}
The following theorem asserts that there exists a formal solution for arbitrary
in\-fi\-ni\-te\-ly dif\-fe\-ren\-ti\-able Cauchy-like data. It also provides its
constructive representation assuming $\hbar \neq 0$.

\begin{theorem}\label{exist}
Let $\tau (x,{\bf t})=f(x)\hat \tau (x+t_1, t_2, \ldots )$ be a tau-function
of the $\hbar$-KP hierarchy with respect to
the variables $t_j$,
with $\tau (x, {\bf 0})$ being an infinitely differentiable function of $x$.
Then the coefficients of the series
\beq\label{Hir105}
\tau (x; {\bf t})=\sum_{\lambda}c_\lambda (x)s_\lambda ({\bf t}/\hbar )
\eeq
are connected by the relations
\beq\label{Hir104}
c_\lambda (x)=(c_0(x))^{1-\ell (\lambda )}
\det_{1\leq i,j\leq \ell (\lambda )}
\left [ \sum_{k=0}^{j-1} (-\hbar )^{k} \left ( \!\!
\begin{array}{c} j\! -\! 1\\ k \end{array}\!\! \right ) \p_x^k
c_{\lambda_i -i+j-k }(x)\right ]
\eeq
where $\left ( \!\!
\begin{array}{c} j\! -\! 1\\ k \end{array}\!\! \right )
=\displaystyle{\frac{(j-1)!}{k! \, (j\! -\! 1\! -\! k)!}}$ is the binomial coefficient,
$c_0(x)=c_{\emptyset}(x)$, $c_k(x)=c_{(k)}(x)$ and
$c_1= \p_x c_0 -c_0\p_x \log f$.

Conversely,
let $\hbar \neq 0$ and $c_k(x)$, $k=0,1,2,\ldots $, be arbitrary infinitely differentiable
functions of $x$ (with $c_0(x)$ being not identically $0$);
for any Young diagram $\lambda$ define the coefficients
$c_\lambda (x)$ by (\ref{Hir104}).
Then the series (\ref{Hir105})
is a formal solution to the $\hbar$-KP hierarchy
($\hbar \neq 0$) with the Cauchy-like data
$$
\tau (x;{\bf 0})= c_0(x), \quad
\p_k^{\hbar}\tau (x;{\bf t})\Bigm |_{{\bf t}=0}=\frac{k}{\hbar} \, c_k(x), \quad
k\geq 1.
$$
\end{theorem}

\noindent
{\it Proof.}
To prove the first part of the theorem,
we note that equations (\ref{Hir3}) for the tau-function $\tau (x;{\bf t})$
can be written in the form
\beq\label{Hir3a}
\prod_{1\leq a<b\leq m}\! (z_b-z_a) \cdot
\tau ^{[z_1, \ldots , z_m]}(x;{\bf t})\tau ^{m-1}(x;{\bf t})=\det_{1\leq i,j\leq m}
\Bigl ((z_i \! -\! \hbar \p_x )^{j-1}\tau ^{[z_i]}(x;{\bf t})\Bigr )
\eeq
(as is easy to see, derivatives of $f$ coming from
$\p_x^l(f\tau )$ cancel in the determinant
in the right hand side after taking proper linear combinations of columns).
The idea is to substitute (\ref{Hir105}) into this equation
and equate the coefficients of the $z_i$-expansion of both sides at ${\bf t}=0$.
Let us rewrite (\ref{Hir3a}) in the form convenient for expanding in powers
of $z_i^{-1}$:
\beq\label{Hir106}
\tau ^{[z_1, \ldots , z_m]}(x;{\bf t})=
\frac{\det \Bigl ( z_i^{j-m}(1-\hbar z_i^{-1}\p_x )^{j-1}
\tau ^{[z_i]}(x;{\bf t})
\Bigr )}{\det \Bigl ( z_i^{j-m}\Bigr ) \, \tau^{m-1}(x;{\bf t})}\,, \quad m\geq 2.
\eeq
It follows from the Cauchy-Littlewood identity that
the expansion of the left hand side is
\beq\label{Hir107}
e^{\hbar (D(z_1)+\ldots + D(z_m))} \tau (x; {\bf t})=
\sum_{\lambda} s_\lambda (\{z_i^{-1}\})s_\lambda (\hbar \tilde \p )
\tau (x; {\bf t})
\eeq
(here only terms with $\ell (\lambda )\leq m$ are non-zero).
The numerator of the right hand side is
$\displaystyle{\det _{1\leq i,j\leq m}\Bigl ( \sum_{l\geq 0}Z_{il}M_{lj}\Bigr )}$, where
$Z_{il}=z_i^{-l}$ is the rectangular semi-infinite $m\! \times \! \ZZ_{\geq 0}$
matrix and
$$
M_{lj}=\sum_{a=0}^{j-1}(-\hbar )^a
\left (\!\! \begin{array}{c}j\! -\! 1\\ a\end{array}\! \! \right ) \p_x^a
h_{l+j-a-m}(\hbar \tilde \p )\tau (x; {\bf t})
$$
is the rectangular semi-infinite $\ZZ_{\geq 0} \! \times \!  m$
matrix (we put $c_{k}=0$ at $k<0$). Application of the Cauchy-Binet formula
gives
$$
\det _{1\leq i,j\leq m}\Bigl ( \sum_{l\geq 0}Z_{il}M_{lj}\Bigr )=
\sum_{l_1>l_2>\ldots > l_m \geq 0}
\det_{ij}\Bigl ( z_i^{-l_j}\Bigr )\cdot
\det_{jk} M_{l_j k}
$$
Setting $l_j=m+\lambda_j -j$, we can represent the multiple sum
as summation over Young
diagrams with rows $\lambda_i$. Using (\ref{id2}),
we see that the expansion of the right hand side
is
$$
\tau (x, {\bf t})^{1-m}\sum_{\lambda}s_{\lambda}(\{z_i^{-1}\})
\det_{ij}\left [ \sum_{a=0}^{j-1}(-\hbar )^a
\left (\!\! \begin{array}{c}j\! -\! 1\\ a\end{array}\!\! \right ) \p_x^a
h_{\lambda_i -i +j-a}(\hbar \tilde \p )\tau (x; {\bf t})
\right ]
$$
Comparing with (\ref{Hir107}),
it remains to put ${\bf t}={\bf 0}$ and use the orthonormality
of the Schur functions.

Now let us prove that the series (\ref{Hir105}) with
$c_{\lambda}$ built from arbitrary $c_k$ according to
(\ref{Hir104}) solves the Hirota
equations (\ref{Hir1}) or (\ref{Hir2}). First we repeat
the above derivation in reverse order to prove that
equations (\ref{Hir3a}) hold for any
$z_1, \ldots , z_m$ and $m\geq 2$ at ${\bf t}=0$. The next step is to
deduce that this set of relations for ``initial values'' of
$\tau ^{[z_1, \ldots , z_m]}$ is equivalent to
{\small $$
(z_2\! -\! z_1) \tau ^{[z_1, z_2, z_3, \ldots , z_m]}(x;{\bf 0})
\tau ^{[z_3, \ldots , z_m]}(x;{\bf 0})=
\left | \begin{array}{ll} \tau^{[z_1, z_3, \ldots , z_m]}(x;{\bf 0})&
(z_1\! -\! \hbar \p_x)\tau^{[z_1, z_3, \ldots , z_m]}(x;{\bf 0})\\ \\
\tau^{[z_2, z_3, \ldots , z_m]}(x;{\bf 0})&
(z_2\! -\! \hbar \p_x)\tau^{[z_2, z_3, \ldots , z_m]}(x;{\bf 0}) \end{array}\right |
$$
}
(which is the Jacobi identity for the matrix
$
N_{ij}=(z_i -\p_x)^{j-1} \tau^{[z_i]}(x;{\bf 0}),
$
see appendix A for details). In its turn, this latter relation, being valid
for all $z_1, \ldots , z_m$ and $m\geq 2$, is equivalent to
\beq\label{Hir108}
(z_2-z_1)\tau ^{[z_1, z_2]}(x;{\bf t}) \, \tau (x;{\bf t})=\left |
\begin{array}{ll} \tau ^{[z_1]}(x;{\bf t}) &
(z_1\! - \! \hbar \p_x )\tau ^{[z_1]}(x;{\bf t})
\\ & \\
\tau ^{[z_2]}(x;{\bf t}) & (z_2\! - \! \hbar \p_x )\tau ^{[z_2]}(x;{\bf t}) \end{array}
\right |
\eeq
for any ${\bf t}$.
This can be formally justified by noticing that if $e^{\hbar (
D(z_1)+\ldots +D(z_k))}{\cal F}({\bf t})\Bigm |_{{\bf t}=0}=0$
for some ${\cal F}({\bf t})$
for all
$z_1, \ldots , z_k$ and $k\geq 0$, then
$\sum_{\lambda}s_{\lambda}(\{z_i^{-1}\})
s_{\lambda}(\tilde \p ){\cal F}({\bf t})\Bigm |_{{\bf t}=0}=0$,
whence $s_{\lambda}(\tilde \p ){\cal F}({\bf t})
\Bigm |_{{\bf t}=0}=0$ for all $\lambda$ which means
that ${\cal F}({\bf t})\equiv 0$ as a formal series.
Multiplying both sides of equation (\ref{Hir108}) by
$\displaystyle{\frac{\tau^{[z_3]}}{\tau^{[z_1]}\tau^{[z_2]}}}$ and summing the
equations obtained in this way for all cyclic permutations of
$\{z_1, z_2, z_3\}$, we get the Hirota equation (\ref{Hir1}) which, by
Proposition \ref{taudet}, is equivalent to (\ref{Hir2}). The latter
equation has the form (\ref{Hir108}) with $\p_x \to \p_1$, so the assertion
is proved.

It remains to show that the tau-function represented by the series
(\ref{Hir105}) does have the form
$f(x)\hat \tau (x+t_1; t_2, \ldots )$, i.e., it essentially depends on $x+t_1$.
For this we notice that the above argument implies that
$$
(\p_x -\p_1)\Bigl ( \log \tau (x; {\bf t}+\hbar
[z^{-1}])-\log \tau (x; {\bf t})\Bigr )=0
$$
for all ${\bf t}$, $z$.
From this it then follows that $\varphi :=(\p_x -\p_1)\log \tau (x; {\bf t})$
does not depend on ${\bf t}$. Indeed, we have
$$
0=\varphi ({\bf t}+\hbar [z^{-1}])-\varphi ({\bf t})=
\hbar \p_1 \varphi \, z^{-1} +\frac{1}{2}(\hbar^2 \p_1^2\varphi +\hbar
\p_2\varphi )z^{-2}+ \ldots
$$
from which we can conclude that $\p_1 \varphi = \p_2 \varphi =\ldots =0$.
Therefore, $(\p_x -\p_1)\log \tau (x; {\bf t})=\varphi (x)$. A simple calculation
at ${\bf t}={\bf 0}$ shows that $\displaystyle{\varphi (x)=\p_x \log c_0(x)-
\frac{c_1(x)}{c_0(x)}}$.
\square
\begin{remark}
Putting $c_1(x)=\p_x c_0(x)$, one obtains tau-functions
$\tau (x+t_1, t_2, \ldots )$ whose $t_1$-evolution is
equivalent to the shift of $x$.
\end{remark}
\begin{remark}
Theorem \ref{exist} has been basically proven in \cite{ALTZ14}, where
equation (\ref{Hir104}) (with $\hbar =1$) has appeared
as a functional equation for commuting transfer matrices
of the quantum Gaudin model.
\end{remark}

\vspace{2ex}


\section{The $\hbar$-KP hierarchy for the $F$-function}

\subsection{Hirota equations for the $F$-function}

For many applications in physics and mathematics one needs to
deal with logarithm of the tau-function rather than with the tau-function itself.
It is possible to represent the Hirota equations as equations for the logarithm.
Let us introduce the $F$-function as
\beq\label{Hir7}
F(x; {\bf t})=
\hbar^2 \log \tau (x;{\bf t}).
\eeq
Representing shifts of the arguments in (\ref{Hir1})
as action of exponential of the differential operators
$D(z_i)$ to the function $F$, we can rewrite (\ref{Hir1}) in
terms of $F$:
\beq\label{Hir4a}
\begin{array}{l}
(z_1-z_2) \exp \! \left [ \hbar^{-2}
e^{\hbar (D(z_1)+D(z_2))}F\right ]
\exp \! \left [\hbar^{-2} e^{\hbar D(z_3)}F\right ]
\\ \\
\hspace{1cm}
+\, (z_2-z_3) \exp \! \left [ \hbar^{-2}
e^{\hbar (D(z_2)+D(z_3))}F\right ]
\exp \! \left [\hbar^{-2} e^{\hbar D(z_1)}F\right ]
\\ \\
\hspace{2cm}
+\, (z_3-z_1) \exp \! \left [ \hbar^{-2}
e^{\hbar (D(z_1)+D(z_3))}F\right ]
\exp \! \left [\hbar^{-2} e^{\hbar D(z_2)}F\right ]=0.
\end{array}
\eeq

One can represent (\ref{Hir4a}) in a more suggestive form
of a non-linear difference equation.
Let us introduce the difference operator
\beq\label{Delta1}
\Delta (z) =\frac{e^{\hbar D(z)}-1}{\hbar}\,.
\eeq
It can be easily checked that equation (\ref{Hir4a}) acquires the form
\beq\label{Delta3}
(z_1-z_2)
e^{\Delta (z_1)\Delta (z_2)F}+
(z_2-z_3)
e^{\Delta (z_2)\Delta (z_3)F}+
(z_3-z_1)
e^{\Delta (z_3)\Delta (z_1)F}=0.
\eeq
Tending $z_3\to\infty$ we get the differential Fay identity (\ref{Hir2})
in terms of $F$:
\beq\label{KP2}
e^{\Delta (z_1)\Delta (z_2)F}=1-
\frac{\Delta (z_1)\p_1 F -\Delta (z_2)\p_1 F}{z_1-z_2}\,.
\eeq
For $F(x;{\bf t})=\phi (x)+\hat F(x+t_1, t_2, t_3, \ldots )$ we can also write this as
\beq\label{KP2a}
e^{\Delta (z_1)\Delta (z_2)F}=1-
\frac{\Delta (z_1)\p_x F -\Delta (z_2)\p_x F}{z_1-z_2}\,.
\eeq

In the form (\ref{Delta3}) the equation is suitable for the $\hbar$-expansion
as $\hbar \to 0$.
The $\hbar$-expansion of the $\hbar$-KP hierarchy was investigated in
\cite{TakTak2,TakTak3} (see also \cite{TakTak4}).
An important class of solutions is distinguished by the assumption that
the function $F$ has a regular
$\hbar$-expansion:
\beq\label{Hir8}
F= F^{(0)} +\hbar F^{(1)} +\hbar^2 F^{(2)} + O(\hbar^3).
\eeq
The case $\hbar =0$ is called the zero
dispersion (or dispersionless) limit.
In this limit
we have $\lim_{\hbar \to 0}\Delta (z)=D(z)$ and equation (\ref{KP2}) becomes
the familiar dispersionless Hirota equation for $F^{(0)}$.


\subsection{The $\hbar$-KP hierarchy in an unfolded form}

Another useful representation of the KP hierarchy is an ``unfolded'' form suggested in \cite{DN,N01}. The notation $\p :=\p_x$,
$F_k:=\p_k F$,  $F_{kl}:=\p_k \p_lF$, etc is convenient. The following theorem can be proven in the same way as \cite[Theorem 2]{N01}, with minor modifications keeping track of the $\hbar$-dependence.
\begin{theorem}\label{unfolded1}
There exist universal rational coefficients $R_{ij}^n\left (
\begin{array}{lll}s_1 & \ldots & s_n \\ r_1 & \ldots & r_n \end{array}\right )$
such that
\beq\label{unf1}
F_{kl}=\sum_{n\geq 1}\sum \hbar^{\sum_i (r_i-1)}
R_{kl}^n\left (
\begin{array}{lll}s_1 & \ldots & s_n \\ r_1 & \ldots & r_n \end{array}\right )
\p^{r_1}F_{s_1}\ldots \p^{r_n}F_{s_n}
\eeq
Here the second sums are taken over all matrices
$\left (
\begin{array}{lll}s_1 & \ldots & s_n \\ r_1 & \ldots & r_n \end{array}\right )$ such that
$s_i, r_i \geq 1$ and $\displaystyle{\sum_{i=1}^n (s_i+r_i)=k+l}$.
\end{theorem}
Equations (\ref{unf1}) represent the KP hierarchy in the unfolded form.
They express $F_{kl}$
with $k,l>1$ as certain universal polynomials
of $\p^rF_{j}$'s with $r,j\geq 1$. The coefficients of these polynomials are
fixed rational numbers which do not depend on a particular solution.
\begin{corollary}\label{unfolded2}
For any integer $m\geq 2$
there exist universal rational coefficients\\ $R_{k_1\ldots  k_m}^n\left (
\begin{array}{lll}s_1 & \ldots & s_n \\ r_1 & \ldots & r_n \end{array}\right )$
such that
\beq\label{unf1a}
F_{k_1 \ldots k_m}=\sum_{n\geq 1}\sum \hbar^{\sum_i (r_i-1)+2-m}
R_{k_1 \ldots k_m}^n\left (
\begin{array}{lll}s_1 & \ldots & s_n \\ r_1 & \ldots & r_n \end{array}\right )
\p^{r_1}F_{s_1}\ldots \p^{r_n}F_{s_n}
\eeq
The second sums are taken over all matrices
$\left (
\begin{array}{lll}s_1 & \ldots & s_n \\ r_1 & \ldots & r_n \end{array}\right )$ such that
$s_i, r_i \geq 1$ with the conditions
$\displaystyle{\sum_{i=1}^n (s_i+r_i)=\sum_{i=1}^n k_i}$,
$\displaystyle{\sum_{i=1}^n r_i\geq n+m-2}$.
\end{corollary}
{\it Proof.} The proof consists in induction on $m$ using the Leibniz rule.
\square

\noindent
It follows from the Corollary that the quantities $F_{j}$, $j\geq 1$ restricted to zero values
of $t_k$ with $k\geq 2$ define the solution up to a constant. In this way, one produces
a unique formal Taylor series
in $t_j$'s which reconstructs a formal solution to the KP hierarchy from the Cauchy data
$f_i(x):= \p_i F(x;{\bf t})\Bigm |_{t_1=t_2=\ldots =0}$. In general,
the functions $f_i$ can be Taylor series in $\hbar$.

The following theorem can be proven in the same way as Theorem 4.2 from
\cite{N03a}.
\begin{theorem} If $\displaystyle{\sum_{i=1}^n (r_i-1)-m\equiv 1 \;
(\mbox{{\rm mod} $2$})}$, then
$R_{k_1\ldots  k_m}^n\left (
\begin{array}{lll}s_1 & \ldots & s_n \\ r_1 & \ldots & r_n \end{array}\right )=0$.
\end{theorem}
We see that equations (\ref{unf1a}) contain even powers of $\hbar$ only.
Therefore, if the Cauchy data $f_i(x)$ are series
in $\hbar^2$, then so is the formal solution. In particular,
the formal solution is expanded in only even powers of $\hbar$ if the Cauchy data
are $\hbar$-independent.

At $\hbar = 0$ only the terms with $r_i=1$ survive in (\ref{unf1})
and these equations yield the
polynomial expressions of $F_{kl}^{(0)}$ with $k,l>1$ through $F_{1j}^{(0)}$'s.
In this case there is a relatively simple recurrence formula for the coefficients, see \cite{N01}.
In general any explicit combinatorial  description of the coefficients
$R_{k_1\ldots k_m}^n$ is not available.

In the rest part of the paper
we show that some closed description of expansion coefficients
can be achieved for non-zero $\hbar$, too,
by a re-summation
of the Taylor series and passing to another basis in the space of polynomials in $t_k$'s.
We will exploit the formal similarity of the Hirota equation (\ref{KP2})
with its dispersionless limit.

\vspace{2ex}


\subsection{The ``$\hbar$-deformed partial derivatives'' $\partial^{\hbar}_i$}

Let us recall the definition of the
``$\hbar$-deformed partial derivatives'' $\partial^{\hbar}_i$ introduced
in (\ref{dhbar}):
\beq\label{dhbar1}
\p^{\hbar}_{k}= \hbar^{-1}k h_k (\hbar \tilde \p ),
\quad \begin{array}{l}
\tilde \p \equiv \bigl \{ \p_1, \frac{1}{2} \p_2,
\frac{1}{3} \p_3, \ldots \bigr \}\end{array}
\eeq
The first few are: $\p^{\hbar}_1 =\p_1$,
$\,\, \p^{\hbar}_2 =\p_2 +\hbar \p_1^2$,
$\,\,
\p^{\hbar}_3 =\p_3 +\frac{3}{2} \hbar \p_1 \p_2 +\frac{1}{2}\hbar^2 \p_1^3$.
It is convenient to put $\p^{\hbar}_0=1$.
Here we discuss these operators in some detail.

The general explicit formula is
\beq\label{Delta103}
\p^{\hbar}_k =\sum_{l=1}^{k}\frac{\hbar ^{l-1}k}{l!}
\! \sum_{{k_1,
\ldots , k_l \geq 1}\atop{k_1+\ldots +k_l=k}}\!\!
\frac{\p_{k_1}\ldots \p_{k_l}}{k_1 \, \ldots \, k_l}\,.
\eeq
There is also a determinant representation which
can be extracted from \cite{Macdonald}:
\beq\label{Delta102}
\p^{\hbar}_n =\frac{1}{(n-1)!}
\left |
\begin{array}{cccccccc}
\p_1 & -1 & 0& 0& & \ldots & & 0
\\
\p_2 &\hbar \p_1 & -2& 0& & \ldots & & 0
\\
\p_3 & \hbar \p_2 & \hbar \p_1 & -3& & \ldots & & 0
\\
\ldots & \ldots & \ldots & \ldots & & \ldots & & \ldots
\\
\p_{n\! -\! 1} & \hbar \p_{n\! -\! 2} & \hbar \p_{n\! -\! 3} &
\hbar \p_{n\! -\! 4}  & & \ldots & &
-(n\! -\! 1)
\\
\p_n & \hbar \p_{n\! -\! 1} & \hbar \p_{n\! -\! 2} &
\hbar \p_{n\! -\! 3} & & \ldots & & \hbar \p_1
\end{array}
\right |.
\eeq

The operators $\p^{\hbar}_k$  have rather special properties. In particular,
they satisfy the following generalized Leibniz rule.

\begin{lemma}\label{Leibniz-lemma}
The differential operators $\p^{\hbar}_k$
satisfy the generalized Leibniz rule:
\beq\label{Leibniz}
\p^{\hbar }_k \prod_{i=1}^n f_i=
\sum_{{k_1, \ldots , k_n\geq 0}\atop{k_1+\ldots +k_n=k}}
\frac{\hbar ^{\nu (k_1,\dots,k_n)-1}k}{[k_1 \ldots k_n]}\,
\p_{k_1}^{\hbar}\! f_{1} \, \p_{k_2}^{\hbar}\! f_{2} \, \ldots \,
\p_{k_n}^{\hbar}\! f_{n}
\eeq
where $\nu (k_1,\dots,k_n)$ is the number of nonzero $k_i$'s
in the sequence $k_1, k_2, \ldots , k_n$ \\ and
$[k_1 \ldots k_n]:=\prod_{i=1}^{n}\max\{k_i,1\}$ is their
product.
\end{lemma}

\noindent
{\it Proof.}
Expanding the both sides of the obvious identity
$\displaystyle{
e^{\hbar D(z)}\prod_{i=1}^n f_i =
\prod_{i=1}^{n}\bigl (e^{\hbar D(z)}f_i \bigr )}
$
in powers of $z$, we get:
$$
\sum_{k\geq 0} z^{-k} {\sf h}_k \prod_{i=1}^n f_i =
\sum_{k_1, \ldots , k_n \geq 0}
(z^{-k_1}{\sf h}_{k_1} f_1 )\ldots (z^{-k_n}{\sf h}_{k_n} f_n ),
$$
where we have put ${\sf h}_k:=h_k (\hbar \tilde \p )$ for brevity.
Equating coefficients in front of powers of $z$ in the both sides,
we obtain the identity\footnote{This means that
the sequence of operators ${\sf h}_0, {\sf h}_1, {\sf h}_2, \ldots$
is a Hasse-Schmidt derivation, see, e.g., \cite{Matsumura}.}
$$
{\sf h}_k \prod_{i=1}^n f_i =\sum_{{k_1, \ldots , k_n \geq 0}\atop{k_1+\ldots +k_n=k}}
({\sf h}_{k_1}f_1 ) \ldots ({\sf h}_{k_n}f_n )
$$
Being rewritten
in terms of the operators $\p_{k}^{\hbar}$, it coincides with (\ref{Leibniz}).
\square

\noindent
In the limit $\hbar \to 0$
(\ref{Leibniz}) becomes the usual Leibniz rule for the partial derivatives $\p_k$.

\vspace{2ex}


\subsection{The $\hbar$-KP hierarchy in terms of $\partial_i^{\hbar}$}

Following \cite{N}, we define the combinatorial constants
$\tilde{P}_{ij}(s_1,\dots,s_m)$ to be
the number of sequences of positive integers
$(i_1,\dots,i_m)$, $(j_1,\dots,j_m)$ such that $i_1+\ldots+i_m=i$, $j_1+\ldots+j_m=j$ and $s_k=i_k+j_k-1$. Put
$$P_{ij}(s_1, \, \ldots , s_m)=
\frac{(-1)^{m+1}ij}{m\, s_1 \ldots s_m }\,
\tilde{P}_{ij}(s_1, \, \ldots , s_m ).$$

\begin{lemma}\label{ld2}
The $\hbar$-KP
hierarchy (\ref{KP2}) is equivalent to the system of equations
\beq\label{f2a}
\p_i^{\hbar}\p_j^{\hbar}F=\sum_{m\geq 1}
\!\! \sum_{{s_1, \ldots , s_m \geq 1}\atop{s_1+\ldots +
s_m=i+j-m}}\!P_{ij}(s_1, \, \ldots , s_m )\,
\p \p_{s_1}^{\hbar} F\, \ldots \, \p \p_{s_m}^{\hbar} F
\eeq
for the function $F(x;\textbf{t})$.
\end{lemma}

\noindent
{\it Proof.} Since
$\displaystyle{\Delta (z) =\sum_{k\geq 1}\frac{z^{-k}}{k}\, \p_k^{\hbar}}$,
the Hirota equation (\ref{KP2}) yields
$$
e^{\Delta(z_1)\Delta(z_2)F}=1+z_1^{-1}z_2^{-1}
\sum\limits_{j=1}^\infty \frac 1j\frac{(z_1^{-j}-z_2^{-j})}{(z_1^{-1}-z_2^{-1})}
\, \partial \partial^{\hbar}_jF
$$
$$
=1+z_1^{-1}z_2^{-1}\sum\limits_{j=1}^\infty\frac 1j
\Bigl (\sum_{{ s+v=j-1}\atop{s, v\geq 0}}
z_1^{-s}z_2^{-v}\Bigr )\, \partial \partial^{\hbar}_j F=
1+\sum\limits_{j=1}^\infty\frac 1j
\Bigl (\! \sum_{{ s+v=j+1}\atop{s,v\geq 1}}
z_1^{-s}z_2^{-v}\Bigr )\,
\partial \partial^{\hbar}_j F.
$$
Therefore,
$$\Delta(z_1)\Delta(z_2)F=\sum\limits_{m=1}^\infty\frac{(-1)^{m+1}}{m}
\left (\sum\limits_{n=1}^\infty
\Bigl (\sum_{{s+v=n+1}\atop{s,v\geq 1}}\! \! z_1^{-s}z_2^{-v}\Bigr )
\frac 1n\, \partial \partial^{\hbar}_n F\right )^m$$
$$=\sum\limits_{j=1}^\infty\frac{(-1)^{m+1}}{m}
\sum\limits_{i,j\geq 1} z_1^{-i}z_2^{-j}
\sum\limits_{{\tiny \begin{array}{c}i_1+\ldots +i_m=i \\
j_1+\ldots +j_m=j \\ i_k, j_k\geq 1\end{array}}}
\frac{\partial \partial^{\hbar}_{i_1+j_1-1}F}{i_1+j_1-1}
\dots \frac{\partial \partial^{\hbar}_{i_m+j_m-1}F}{i_m+j_m-1} $$
that is
$$\partial^{\hbar}_i\partial^{\hbar}_j F= \sum\limits_{m=1}^\infty
\sum\limits_{k_1+\ldots +k_m=i+j-m}\frac {(-1)^{m+1}ij}{m\, k_1 \ldots k_m}
\, \tilde{P}_{ij}(k_1,\ldots ,k_m)\,
\partial \partial^{\hbar}_{k_1} F \, \ldots \,
\partial \partial^{\hbar}_{k_m} F.$$
This proves that (\ref{KP2}) implies (\ref{f2a}).
Performing the calculation
in reverse order, we find that (\ref{f2a}) implies (\ref{KP2}).
\square

\noindent
In particular, we have $\p_1 \p_j^\hbar F=\p \p_j^\hbar F$.

\vspace{3ex}

Let $K_l(l^1,\dots,l^r)$ be the number of partitions of a set of $l$ elements into ordered groups of $l^1,\dots,l^r$ elements.
Define constants $P^{\hbar}_{i_1\ldots i_k}
\bpmatrix  s_1\ldots s_m \\l_1\ldots l_m \epmatrix$ by
the following recurrence relations:

\begin{itemize}\label{Prek}
\item[1)]$P^{\hbar}_{i_1,i_2}\bpmatrix  s_1\ldots s_m
\\1\ldots 1\epmatrix=P_{i_1i_2}(s_1, \, \ldots \, ,s_m)$ and  $P^{\hbar}_{i_1,i_2}\bpmatrix s_1\ldots
s_m \\l_1\ldots l_m \epmatrix  =0$, if $\prod\limits_{j=1}^ml_j>1$.
\item[2)]
$\displaystyle{P^{\hbar}_{i_1\dots i_r}
\bpmatrix x_1\ldots x_v \\y_1\ldots y_v\epmatrix   =
\sum P^{\hbar}_{i_1\dots i_{r-1}}\bpmatrix s_1\ldots s_m $
$l_1\ldots l_m\epmatrix  \frac{\hbar ^{\nu (k_1,
\ldots ,k_m)-1}i_r} {[k_1 \ldots k_m]}}\\
\displaystyle{\times \,K_{l_1}(l_1^1,\ldots ,l_1^{n_1}) P^{\hbar}_{s_1k_1}(s_1^1 \, \ldots \, s_{n_1}^1)  \ldots K_{l_m}(l_m^1,\ldots ,l_m^{n_m})}P^{\hbar}_{s_mk_m}(s_1^m \, \ldots \, s_{n_{m}}^m),$
\end{itemize}
where  the summation is carried over all sets
of integer numbers $m>0$, $s_i>0$, $s_j^i>0$, $k_i\geq0$, $l_i>0$, $l_j^i\geq0$ such that
$$(x_1\ldots x_v)=(s_1^1, \, \ldots \, s_{n_1}^1,s_1^2,
\ldots ,s_{n_2}^2,\ldots, s_1^m, \ldots,s^m_{n_m}),
\quad s_i= \sum_{j=1}^{n_i} s^i_j;$$
$$(y_1\ldots y_v)=(l^1_1+1,\ldots,l^1_{n_1}+1,l_1^2+1, \ldots l_{n_2}^2+1,\ldots,l^m_1+1, \ldots,l^m_{n_m}+1), \quad l_i=\sum_{j=1}^{n_i} l^i_j$$
$$\sum_{i=1}^m (s_i+l_i)= \sum_{j=1}^{r-1}i_j, \quad \sum_{i=1}^m k_i=i_r, \quad \sum_{i=1}^{n_j}  s_i^j= k_j+s_j.$$

\vspace{2ex}

Lemma \ref{ld2} and Lemma \ref{Leibniz-lemma} imply

\
\begin{theorem}\label{td2}
The $\hbar$-KP hierarchy (\ref{KP2}) is equivalent to the system of equations
\beq\label{pppp}
\p_{i_1}^{\hbar}\p_{i_2}^{\hbar}\ldots\p_{i_r}^{\hbar} F= \sum_{m\geq1}\sum_{{\tiny \begin{array}{c}
s_1+l_1+\ldots+s_m+l_m\\ =i_1+\dots+ i_{r} \\
1\leq s_i; \, 1\leq l_i\leq r-1\end{array}}}P^{\hbar}_{i_1\ldots i_{r}}
\bpmatrix s_1\ldots s_m \\l_1\ldots l_m\epmatrix   \p^{l_1} \p_{s_1}^{\hbar}F\, \ldots \,\p^{l_m}\p_{s_m }^{\hbar}  F
\eeq
(here $r\geq 2$).
\end{theorem}

\noindent
{\it Proof.} We use induction in $r$. At $r=2$ Lemma \ref{ld2} gives
\beq
\p_{i_1}^{\hbar}\p_{i_2}^{\hbar}F= \sum_{m\geq1}\sum_{{\tiny \begin{array}{c}
s_1+1+\ldots+s_m+1 \\ =i_1+i_2 \\
1\leq s_i \end{array}}} P^{\hbar}_{i_1i_2}
\bpmatrix s_1\ldots s_m \\1\ldots 1\epmatrix   \p \p_{s_1}^{\hbar}F\,
\ldots \,\p \p_{s_m }^{\hbar}  F.
\eeq
Then we calculate $\p_{i_1}^{\hbar}\p_{i_2}^{\hbar}\ldots\p_{i_r}^{\hbar} F=
\p_{i_r}^{\hbar}(\p_{i_1}^{\hbar}\p_{i_2}^{\hbar}
\ldots\p_{i_{r-1}}^{\hbar} F)$ assuming that
$\p_{i_1}^{\hbar}\p_{i_2}^{\hbar}
\ldots\p_{i_{r-1}}^{\hbar} F$ is given by the assertion of the theorem.
We have, using the generalized Leibniz rule (\ref{Leibniz}):
$$\p_{i_1}^{\hbar}\p_{i_2}^{\hbar}\ldots\p_{i_r}^{\hbar} F=
\p_{i_r}^{\hbar}(\p_{i_1}^{\hbar}\p_{i_2}^{\hbar}
\ldots\p_{i_{r-1}}^{\hbar} F)$$
$$=\, \sum_{m\geq1}\sum_{{\tiny
\begin{array}{c}s_1+l_1+\ldots +s_m+l_m
\\ =i_1+\ldots +i_{r-1} \\ 1\leq s_i; \, 1\leq l_i\leq r-2 \end{array}}}
P^{\hbar}_{i_1\ldots i_{r-1}}\bpmatrix s_1\ldots s_m \\l_1
\ldots l_m\epmatrix  \partial_{i_r}^{\hbar}( \p^{l_1} \p_{s_1}^{\hbar}F\,
\ldots \,\p^{l_m}\p_{s_m }^{\hbar}  F)$$
$$=\,
\sum_{m\geq1}\sum_{{\tiny \begin{array}{c}
s_1+l_1+\ldots+s_m+l_m \\= i_1+\ldots+i_{r-1} \\1\leq s_i; \, 1\leq l_i\leq r-2 \end{array}}}P^{\hbar}_{i_1\ldots i_{r-1}}\bpmatrix s_1\ldots s_m \\l_1\ldots l_m\epmatrix  \sum_{{k_1, \ldots , k_m\geq 0} \atop{k_1+\ldots +k_m=i_r}}\frac{\hbar ^{\nu (k_1,\dots,k_m)-1}i_r}{[k_1 \ldots k_m]}$$
$$\p^{l_1}\p_{s_1}^{\hbar}\p_{k_1}^{\hbar}F\, \ldots \,\p^{l_m}\p_{s_m }^{\hbar}\p_{k_m }^{\hbar}F$$

$$=\,
\sum_{m\geq1}\sum_{{\tiny
\begin{array}{c}s_1+l_1+\ldots +s_m+l_m
\\= i_1+\ldots +i_{r-1} \\ 1\leq s_i; \,
1\leq l_i\leq r-2\end{array}}}P^{\hbar}_{i_1\ldots i_{r-1}}
\bpmatrix s_1\ldots s_m \\l_1\ldots l_m\epmatrix
\sum_{{k_1, \ldots , k_m\geq 0}\atop{k_1+\ldots +k_m=i_r}}
\frac{\hbar ^{\nu (k_1,\dots,k_m)-1}i_r}{[k_1 \ldots k_m]}$$
$$
\p^{l_1}\Bigl(\sum_{n_1\geq 1}\!\! \sum_{{s_1, \ldots , s_{n_1} \geq 1} \atop{s_1^1+\ldots +s_{n_1}^1+{n_1}=k_1+s_1}}\!
P^{\hbar}_{s_1k_1}(s_1^1\ldots s_{n_1}^1)  \,\p\p_{s_1^1}^{\hbar} F\,
\ldots \, \p\p_{s_{n_1}^1}^{\hbar} F\Bigr) \dots$$
$$\p^{l_m}\Bigl(\sum_{n_m\geq 1}\!\! \sum_{{s^m_1,
\ldots , s^m_{n_m} \geq 1} \atop{s^m_1+\ldots+s^m_{n_m}+{n_m}= k_m+s_m}}\!
P^{\hbar}_{s_mk_m}(s_1^m\ldots s_{n_m}^m)\,\p \p_{s_1^m}^{\hbar}
F\, \ldots \, \p\p_{s_{n_m}^m}^{\hbar} F\Bigr),$$
which is equivalent to (\ref{pppp}).
\square


\section{Formal solution for the $F$-function}

\subsection{The Cauchy-like data}

Similarly to Definition \ref{def-Cauchy-like}, we can introduce the Cauchy-like
data for the $\hbar$-KP hierarchy for $F$: $F(x; {\bf 0})$ and
$\p_k^{\hbar}F(x;{\bf 0}):=\p_k^{\hbar}F(x;{\bf t})\Bigm |_{{\bf t}=0}$.
Let us examine how the Cauchy-like data for $\tau$ and $F$ are connected.
First of all, we have
\beq\label{tauF0}
\tau (x;{\bf 0})=c_0(x)=e^{F(x;{\bf 0})/\hbar ^2}.
\eeq
Next, we use the obvious identity
$\displaystyle{
\hbar \frac{\Delta (z)\tau}{\tau}=e^{\frac{1}{\hbar}\Delta (z)F}-1}
$
which immediately follows from the definition of $\Delta (z)$ \footnote{
It is valid for {\it any} functions $\tau$, $F$ related by
$F=\hbar ^2 \log \tau$ (not necessarily KP solutions).}.
Expanding it in powers of $z$ and comparing the coefficients, we get:
\beq\label{tauF}
\left. \frac{\p_k^\hbar \tau (x; {\bf t})}{\tau (x;{\bf 0})}\right |_{{\bf t}=0}
=\frac{k}{\hbar}\, h_k ({\bf y}), \quad
y_l= \frac{1}{\hbar l}\, \p_l^\hbar F(x; {\bf t})\Bigm |_{{\bf t}=0},
\eeq
or, in terms of the coefficients $c_k(x)$ introduced in Theorem \ref{exist},
\beq\label{tauF1}
\frac{c_k(x)}{c_0(x)}=h_k({\bf y})\,, \quad k=1,2, \ldots
\eeq
Therefore, the Cauchy-like data for $\tau$ are unambiguously determined by
the ones for $F$.
Theorem \ref{exist} implies the following
\begin{corollary}\label{existF} Let $\hbar \neq 0$.
Then for any Cauchy-like data $F(x; {\bf 0})$,
$\p_k^{\hbar}F(x;{\bf 0})$, $k\geq 1$,
there exists a formal solution $F(x;{\bf t})$ to the $\hbar$-KP hierarchy.
\end{corollary}

\subsection{Dual bases $h_{\lambda}$ and $m_{\lambda}$ and the formal solution}

Theorem \ref{td2} gives $h_{\lambda}(\hbar \tilde \p )F (x;{\bf 0})$ in terms
of the Cauchy-like data. Therefore, for the construction of the Taylor series
we need the dual bases
$h_\lambda , m_{\mu}$.
The functions $m_{\mu}({\bf t})$ will be used
in the expansion of the $F$-function instead of monomials
$t_{\lambda}$, according to (\ref{dual2}).

It is not difficult to incorporate the parameter $\hbar$ into the formulas.
For instance, re-scaling ${\bf t}\to {\bf t}/\hbar$ in
the generalized Taylor expansion (\ref{dual2}) with respect to a dual pair
$\left <u_\lambda , v_\mu \right >=\delta_{\lambda \mu}$
and redefining the function $g$ as $g({\bf t}/\hbar )=G({\bf t})$,
we can rewrite it in the form
\beq\label{dual2a}
G({\bf t})=\sum_{\lambda}u_{\lambda}(\hbar \tilde \p )G({\bf t}')
\Bigr |_{{\bf t}'=0} \cdot
v_{\lambda}({\bf t}/\hbar ).
\eeq
Here, $G({\bf t})$ is an arbitrary series in the $t_j$'s with
$\hbar$-dependent coefficients. Applying this formula for the dual pair
$h_{\lambda}$, $m_{\lambda}$ and the function $F(x;{\bf t})$, we can write:
\beq\label{dual2b}
F(x;{\bf t})=\sum_{\lambda}h_{\lambda}(\hbar \tilde \p )F(x;{\bf t}')
\Bigr |_{{\bf t}'=0} \cdot
m_{\lambda}({\bf t}/\hbar ).
\eeq

As the $\hbar$-deformation of the monomial basis $t_{\lambda}$ we
introduce the polynomials
\beq\label{dual14aa}
t_{\lambda}^{\hbar}:=\frac{\sigma (\lambda )}{\rho (\lambda )}\,
\hbar^{\ell (\lambda )}m_{\lambda}({\bf t}/\hbar ).
\eeq
The first few are (see (\ref{dual15a})):
\beq\label{dual15}
\begin{array}{l}
t^{\hbar}_{(1)}=t_1
\\  \\
t^{\hbar}_{(2)}=t_2, \quad
t^{\hbar}_{(1^2)}= t_1^2  -2\hbar t_2
\\  \\
t^{\hbar}_{(3)}=t_3, \quad
t^{\hbar}_{(21)}=t_2 t_1 -\frac{3}{2} \hbar t_3, \quad
t^{\hbar}_{(1^3)} = t_1^3 - 6\hbar t_2 t_1 +6\hbar^2 t_3.
\end{array}
\eeq
Clearly, $t_{\lambda}^{\hbar =0}=t_{\lambda}$.
In these terms,
the Taylor series for the function $F$ can be written as follows:
\beq\label{dual16}
F(x;{\bf t})=\sum_{\lambda}\p_{\lambda}^{\hbar}
\, F(x; {\bf t}')\Bigr |_{{\bf t}'=0}\,
\frac{t_{\lambda}^{\hbar}}{\sigma (\lambda )}\,,
\eeq
where $\p_{\lambda}^{\hbar}:=\p_{\lambda_1}^{\hbar}\p_{\lambda_2}^{\hbar}\ldots
\p_{\lambda_\ell}^{\hbar}$.
All solutions of system (\ref{f2a}) of this form
are called \textit{formal solutions}.
An important corollary of Theorem \ref{td2} is
\begin{corollary}\label{Cau}
Any formal solution $F(x;\textbf{t})$ of (\ref{KP2})
is uniquely defined by the Cauchy-like data $F(x,{\bf 0})$
and $\{\p_{k}^{\hbar} F(x;{\bf 0})|\, k=1,2,\dots\}$.
\end{corollary}

\subsection{$t^{\hbar}_{\lambda}$ in terms of $t_{\lambda}$}

Following \cite{Macdonald},
consider the transition matrix $L_{\lambda \mu}$ from
power sums to the monomial symmetric functions:
$$
p_{\lambda}({\bf t})=\sum_{\mu}L_{\lambda \mu}m_{\mu}({\bf t}).
$$
It has the following combinatorial description.
Let $\lambda$ be a partition of length $\ell =\ell (\lambda )$,
and $f$ be any mapping from the set $\{1, 2, \ldots , \ell\}$
to the set of positive integers. Consider
the infinite sequence
$\{f^{(\lambda )}_1, f^{(\lambda )}_2, \ldots \}$ whose
{\it i}th component is
$$
f^{(\lambda )}_i=\sum_{j: \,\, f(j)=i}\! \lambda_j\quad \quad
\mbox{for each $i\geq 1$}.
$$
\begin{theorem} (\cite{Macdonald})
$L_{\lambda \mu}$ is equal to the number of mappings $f$
such that $f^{(\lambda )}_i =\mu_i$ for each $i\geq 1$.
\end{theorem}

\noindent
It follows from the theorem that $L_{\lambda \mu}$ are non-negative
integer numbers.
Let us regard the partitions of $n=|\lambda |$ as arranged
in the reverse
lexicographical order.
It follows from the theorem that
$L_{\lambda \mu}=0$ unless
$\mu$ precedes $\lambda$. In fact a stronger property
holds: $L_{\lambda \mu}=0$ unless $\mu \geq \lambda$, where the
partial ordering $\geq$ is defined in (\ref{ordering}).
This means that the matrix $L_{\lambda \mu}$ is
strictly lower triangular:
\beq\label{dual13}
p_{\lambda}({\bf t})=\sum_{\mu \geq \lambda}L_{\lambda \mu}m_{\mu}({\bf t}),
\quad
m_{\lambda}({\bf t})=\sum_{\mu \geq \lambda}
(L^{-1})_{\lambda \mu}\, p_{\mu}({\bf t}).
\eeq
(in particular, $L_{\lambda \mu }=(L^{-1})_{\lambda \mu}=0$ if
$\lambda$ and $\mu$ are incomparable with respect to the partial ordering $\geq$).
We can also rewrite (\ref{dual13}) as
\beq\label{dual14}
t_{\lambda}^{\hbar}=\frac{\sigma (\lambda )}{\rho (\lambda )}
\sum_{\mu \geq \lambda} (L^{-1})_{\lambda \mu}\,\rho (\mu )
\hbar^{\ell (\lambda )-\ell (\mu )}t_{\mu}.
\eeq
Let us stress that the sums in (\ref{dual13}), (\ref{dual14}) are
finite: it is implied that $|\mu |=|\lambda |$.

\vspace{2ex}

\subsection{Algebra of $\hbar$-differential operators}

Let $F$ be a solution of the
$\hbar$-KP hierarchy. It follows from
Theorem \ref{td2} that $\{\partial_i^{\hbar}\, | \, i=1,2,\dots\}$ act on the algebra generated by $\{\partial_1^l\partial^{\hbar}_s
F(x;t_1,0,0,\dots)\, | \, l,s=1,2,\dots\}$.
In this subsection we prove that any family of functions of one variable generates a similar commutative algebra of differential operators.
Later we use this algebra for constructing all formal solutions.

Consider an arbitrary set $\textbf{f}=\{f_1,f_2,\dots\}$ of real or complex formal functions of
the variable $x$.
Let $A_{\textbf{f}}$ be the algebra generated by $\{f_i\}$
and their derivatives $\partial^nf_i= \frac{\partial^nf_i}{\partial x^n}$.

Consider a family of linear operators $L_i^{\hbar}=L_i^{\hbar}[\textbf{f}]: A_{\textbf{f}}\rightarrow A_{\textbf{f}}$\,\,,
which are uniquely defined by the following properties:
\begin{itemize}
\item
$L_i^{\hbar}\partial=\partial L_i^{\hbar}$;
\item
$\displaystyle{L_i^{\hbar}(f_j)=\sum_{m\geq 1}
\!\! \sum_{{s_1, \ldots , s_m > 1}
\atop{s_1+\ldots +s_m=i+j}}\!P_{ij}(s_1-1, \, \ldots \, , s_m-1 )\, \partial f_{s_1-1}\, \ldots \, \partial f_{s_m-1}}$;
\item (the generalized Leibniz rule) for any formal functions $g_1,\dots,g_n$,
of $t$ and $\hbar$ it holds
\beq\label{Leibniz2}
L_k^{\hbar}(g_1\ldots g_n)=
\sum_{{k_1, \ldots , k_n\geq 0}\atop{k_1+\ldots +k_n=k}}
\frac{\hbar ^{\nu (k_1,\dots,k_n)-1}k}{[k_1 \ldots k_n]}\,
L_{k_1}^{\hbar}\! g_{1} \, L_{k_2}^{\hbar}\! g_{2} \, \ldots \,L_{k_n}^{\hbar}\! g_{n};
\eeq
\end{itemize}

\noindent
In particular, $L_1^{\hbar}(f_j)=\sum\limits_{m\geq 1}
\!\! \sum\limits_{{s_1, \ldots , s_m > 1}
\atop{s_1+\ldots +s_m=1+j}}\!P_{1j}(s_1-1, \, \ldots \, , s_m-1 )\, \partial f_{s_1-1}\, \ldots \, \partial f_{s_m-1}= \partial f_j$ and $L_j^{\hbar}(f_i)=  L_i^{\hbar}(f_j)$ because $P_{ji}(s_1, \, \ldots \, s_m )=P_{ij}(s_1, \, \ldots \, s_m )$.

The lemma below follows from Lemma \ref{ld2}.
(This is a motivation for introducing the operators $L_i^{\hbar}$.)

\begin{lemma}\label{operators} Let $F$ be the solution to the $\hbar$-KP
hierarchy corresponding to the Cauchy-like data
$F(x,{\bf 0})$ and $\{\p_{k}^{\hbar} F(x,{\bf 0})= f_k(x)| \, k=1,2,\dots\}$.
Then $$L_i^{\hbar}(f_j)(x)=\partial_i^{\hbar}\partial_j^{\hbar} F(x,{\bf 0}).$$
\end{lemma}


\begin{theorem}\label{algebr} The operators $L_i^{\hbar}$ commute with each other.
\end{theorem}
\noindent
{\it Proof.} First assume that
$\hbar \neq 0$. Then it follows from Corollary \ref{existF} that
there is a unique solution $F$ to
the $\hbar$-KP hierarchy with the Cauchy-like data $\textbf{f}$. Hence
$$L_k^{\hbar}L_i^{\hbar}(f_j)= \partial_k^{\hbar}
\partial_i^{\hbar}\partial_j^{\hbar} F(x;{\bf t})=L_i^{\hbar}L_k^{\hbar}(f_j).$$
The operators $L_i^{\hbar}$ are presented in the form of formal series
$L_i^{\hbar}=\sum\limits_{j=0}^{\infty}\hbar^j H_i^j$, where the linear operators $H_i^j=H_i^j[\textbf{f}]: A_{\textbf{f}}\rightarrow A_{\textbf{f}}$\,\, do not
depend on $\hbar$ and $H_i^0=L_i^0$.
The equalities
$L_k^{\hbar}L_i^{\hbar}(f_j)= L_i^{\hbar}L_k^{\hbar}(f_j)$ for all
non-zero $\hbar$ give $H_k^0H_i^0(f_j)= H_i^0H_k^0(f_j)$.
\square
\begin{corollary}\label{commut}
For any $i_1,\dots,i_r$ and $\sigma\in S_r$ the relations
$$
L_{i_1}^{\hbar}\dots L_{i_{r-1}}^{\hbar}(f_{i_r})= L_{i_{\sigma(1)}}^{\hbar}\dots L_{i_{\sigma(r-1)}}^{\hbar}(f_{i_{\sigma(r)}})
$$ and
$$
L_{i_1}^{\hbar}\dots L_{i_{r-1}}^{\hbar}(f_{i_r}) =\sum\limits_{m\geq1}\sum\limits_{{\tiny \begin{array}{c}
s_1+l_1+\ldots+s_m+l_m=\\i_1+\dots+ i_{r} \\
1\leq s_i; \, 1\leq l_i\leq r-1\end{array}}}P^{\hbar}_{i_1\ldots i_{r}}\bpmatrix s_1\ldots s_m \\l_1\ldots l_m\epmatrix   \p^{l_1} f_{s_1}\, \ldots \,\p^{l_m}f_{s_m }
$$
are fulfilled.
\end{corollary}
\noindent
{\it Proof.} The first relation directly follows from Theorem \ref{algebr}.
The second one follows from Lemma \ref{operators} and
Theorem \ref{td2}.
\square


\subsection{Construction of formal solutions}

Given a Young diagram $\lambda=[\lambda_1, \dots \, , \lambda_\ell ]$, we put
$P_{\lambda}^{\hbar} \bpmatrix s_1\ldots s_m
\\l_1\ldots l_m\epmatrix  = P_{\lambda_1\ldots \lambda_r}^{\hbar} \bpmatrix s_1\ldots s_m \\l_1\ldots l_m\epmatrix$.
\begin{theorem}\label{formalDKP}
For any ${\hbar}$ and any family of smooth or formal functions
$$\textbf{f}= \{f_0(x), f_{1}(x), f_{2}(x), \ldots \}$$
there exists a unique solution $F(x;{\bf t})$ of the ${\hbar}$-\textsc{KP}
hierarchy  such that $F(x;{\bf 0})= f_0(x)$ and
$\partial^{\hbar}_k F(x;t_1, t_2, \ldots )\!
\Bigm |_{{\bf t}=0}= f_{k}(x)$. This solution has the form
\beq\label{dual19}
F(x;{\bf t})= f_0(x)+
\sum_{|\lambda|\geq 1}\, \frac{f_{\lambda}^{\hbar}(x)}{\sigma (\lambda )}\,
t_{\lambda}^{\hbar}\,,
\eeq
where $f_{(k)}^{\hbar}(x)=f_k(x)$ and
\beq\label{dual0}
f_{\lambda}^{\hbar}(x)= \sum_{m\geq1}\sum_{{\tiny \begin{array}{c}
s_1+l_1+\ldots+s_m+l_m=|\lambda|\\
1\leq s_i; \, 1\leq l_i\leq\ell(\lambda)-1
\end{array}}} P_{\lambda}^{\hbar}\bpmatrix s_1\ldots s_m \\l_1\ldots l_m\epmatrix
\p^{l_1} f_{s_1}(x)\, \ldots \,\p ^{l_m}f_{s_m}(x)
\eeq
for $\ell(\lambda)>1$.
\end{theorem}

\noindent
{\it Proof.} Let us consider the operators $L^{\hbar}_i$
generated by the family $\{f_{i}|\,i=0,1,2, \ldots\}$.
Given a Young diagram $\lambda=[\lambda_1,\dots,\lambda_r]$,
denote by $[i,j,\lambda]$ the Young diagram with the rows
$\{i,j,\lambda_1,\dots,\lambda_r\}$ (ordered according to their length).
Then, according to Corollary \ref{commut},
$$\partial_{i}^{\hbar}\partial_{j}^{\hbar}F= \sum_{\lambda}\,
\frac{f_{[i,j,\lambda]}(x)}{\sigma ([i,j,\lambda])} \, t_{\lambda}^{\hbar}\,=
\sum_{\lambda}\, \frac{L_i^{\hbar}L_j^{\hbar}f_{\lambda}^{\hbar}(x)}{\sigma (\lambda )}\, t_{\lambda}^{\hbar} = L_i^{\hbar}L_j^{\hbar}\sum_{\lambda}\,
\frac{f_{\lambda}^{\hbar}(x)}{\sigma (\lambda )}\, t_{\lambda}^{\hbar},$$
for $i,j\geq 1$.
Hence, $$\partial_{i}^{\hbar}\partial_{j}^{\hbar}
F(x;t_1,t_2,\dots)\Bigm |_{{\bf t}=0}= L_i^{\hbar}L_j^{\hbar}F(x;{\bf 0})$$
and
$$\partial_{i}^{\hbar}\partial_{j}^{\hbar}F= \sum_{m\geq 1}\!\! \sum_{{s_1, \ldots , s_m > 1}\atop{s_1+\ldots +
s_m=i+j}}\!P_{ij}(s_1-1, \, \ldots , s_m-1 )\,
\p\p_{s_1-1}^{\hbar}F\, \ldots \, \p \p_{s_m-1}^{\hbar}F^{\hbar}$$
Therefore, according to Lemma \ref{ld2}, the function $F$ is a formal solution
of the ${\hbar}$-\textsc{KP} hierarchy. Moreover, according to Corollary \ref{Cau},
the solution is uniquely determined by the functions
$f_{k}(x)$.
\square

\begin{corollary}\label{Cauchy} The Cauchy data are connected with the
Cauchy-like data by
\beq\label{dual20}
\partial_k F(x;{\bf t})\Bigm |_{{\bf t}=0}=k \sum_{|\lambda|=k}
\frac{\kappa_{\lambda}}{\rho (\lambda)} \, f^{\hbar}_{\lambda}(x)\,
\hbar^{\ell (\lambda )-1},
\eeq
where $\kappa_{\lambda}=(L^{-1})_{\lambda (k)}$.
For any family formal functions $\textbf{f}= \{f_0(x), f_{1}(x), f_{2}(x), \ldots \}$ there exists a single solution of $\hbar$-KP hierarchy $F(x;{\bf t})$  with Cauchy data $\textbf{f}$.
\end{corollary}

\noindent
{\it Proof.} The first statement immediately follows from (\ref{dual14}) and (\ref{dual19}). According to Theorem \ref{formalDKP}, the second statement follows from an inversion of  (\ref{dual20}). This inversion is proved by induction in $k$, according to
\beq\label{dual21}
f^{\hbar}_{k}(x) =\partial_k F(x;{\bf t})\Bigm |_{{\bf t}=0}-\, k \!\sum_{|\lambda|=k\atop \ell(\lambda)>1}
\frac{\kappa_{\lambda}}{\rho (\lambda)} \, f^{\hbar}_{\lambda}(x)\,
\hbar^{\ell (\lambda )-1},
\eeq
\square

\section*{Acknowledgments}
The work of S.N. was supported in part by RFBR grant
15-52-50041, and by grant Nsh-5138.2014.1 for support of scientific schools.
The work of A.Z. was supported in part by RFBR grants
14-02-00627, 15-52-50041-YaF, 14-01-90405-Ukr and by grant
Nsh-1500.2014.2 for support of scientific schools.
The financial support from the Government of the Russian Federation
within the framework of the implementation of the 5-100 Programme Roadmap of
the National Research University  Higher School of Economics is acknowledged.

\section*{Appendix A: Proof of Proposition \ref{taudet}}
\addcontentsline{toc}{section}{Appendix A:
Proof of Proposition \ref{taudet}}
\def\theequation{A\arabic{equation}}
\def\theHequation{\theequation}
\setcounter{equation}{0}

Here we give some details of the proof that the tau-function of the
$\hbar$-KP hierarchy satisfies (\ref{Hir3}) for any $m\geq 2$:
\beq\label{Hir3aa}
\prod_{1\leq i<j\leq m}\! (z_j-z_i) \cdot
\tau ^{[z_1, \ldots , z_m]}\tau ^{m-1}=\det_{1\leq j,k\leq m}
\Bigl ((z_j \! -\! \hbar \p_1 )^{k-1}\tau ^{[z_j]}\Bigr )
\eeq
(Proposition \ref{taudet}).
At $m=2$ equation (\ref{Hir3aa}) reads
\beq\label{ap1}
(z_2-z_1)\tau ^{[z_1, z_2]} \tau =\left |
\begin{array}{ll} \tau ^{[z_1]} & (z_1\! - \! \hbar \p_1 )\tau ^{[z_1]}
\\ & \\
\tau ^{[z_2]} & (z_2\! - \! \hbar \p_1 )\tau ^{[z_2]} \end{array}
\right |
\eeq
which is the original equation for the tau-function (\ref{Hir2}).
The rest of the proof is induction in $m$.

Suppose (\ref{Hir3aa}) holds
for any number of points $z_i$ from $2$ to $m-1$. In what follows it is convenient
to use the short-hand notation
$$
\Delta_{1\ldots m}=\prod_{a>b}^m (z_a-z_b), \quad
\tau_{1\ldots m}=\tau \Bigl ( {\bf t}+\hbar [z_1^{-1}]+\ldots +
\hbar [z_m^{-1}]\Bigr ).
$$
After the shift $\displaystyle{{\bf t}\to {\bf t}+\sum_{i=3}^m [z_i^{-1}]}$
equation (\ref{ap1}) acquires the form
\beq\label{ap3}
\Delta_{12}\, \tau_{123\ldots m}\, \tau_{3\ldots m}=\left |
\begin{array}{ll} \tau_{13\ldots m} & (z_1\! - \! \hbar \p_1 )\tau_{13\ldots m}
\\ & \\
\tau_{23\ldots m} & (z_2\! - \! \hbar \p_1 )\tau_{23\ldots m}  \end{array}
\right |.
\eeq
Let $N_{ij}$ be the matrix
$
N_{ij}=(z_i-\hbar \p_1)^{j-1}\tau^{[z_i]}$,  $1\leq i,j\leq m$,
and $\tilde N_{ij}$ be the matrix whose first $m-1$ columns are the same as
for $N_{ij}$ but the last column is different: $\tilde N_{i  \, m}=N_{i \, m\! +\! 1}$.
One can check \footnote{The proof is based on the easily verified identity
$\displaystyle{\sum_{l=1}^{m}\det \Bigl (
z_i^{\delta_{jl}}\! A_{ij}\Bigr )= \Bigl ( \sum_{l=1}^m z_l\Bigr )\det A_{ij}}$
valid for any matrix $A_{ij}$} that
\beq\label{ap2}
\hbar \p_1 \det N_{ij}= \Bigl ( \sum_{l=1}^m z_l\Bigr ) \det N_{ij}-\det \tilde N_{ij}
\eeq
By $N_{ij}\bigl [ a,b\bigr ]$ we denote the matrix $N_{ij}$
with removed row $a$ and column $b$, and by
$N_{ij}\Bigl [ {\small \begin{array}{c} a,b\\ c,d\end{array}}\Bigr ]$
we denote the matrix $N_{ij}$
with removed rows $a,c$ and column $b,d$. By the assumption of the induction we have:
$$
\tau_{3\ldots m}=\frac{\det N_{ij}
\Bigl [ {\small \begin{array}{cc} 1,& \!\! m\! -\! 1\\ 2, &
\!\! m\end{array}}\Bigr ]}{\Delta_{3\ldots m}\, \tau^{m-3}}\,,
\quad
\tau_{13\ldots m}=\frac{\det N_{ij}[2,m]}{\Delta_{13\ldots m}\, \tau^{m-2}}\,,
\quad
\tau_{23\ldots m}=\frac{\det N_{ij}[1,m]}{\Delta_{23\ldots m}\, \tau^{m-2}}\,.
$$
Substituting this into (\ref{ap3}), we get, after using (\ref{ap2}):
$$
\Delta_{12} \, \tau_{1\ldots m}\cdot
\frac{\det N
\Bigl [ {\small \begin{array}{cc} 1,& \!\! m\! -\! 1\\ 2, &
\!\! m\end{array}}\Bigr ]}{\Delta_{3\ldots m}\, \tau^{m-3}}=
\frac{\left | \begin{array}{cc} \det N [2,m] & \det N [2, m\! -\! 1]\\ &\\
\det N [1,m] & \det N [1, m\! -\! 1]\end{array}\right |}{\Delta_{13\ldots m}\,
\Delta_{23\ldots m}\tau^{2m-4}}
$$
Since $\Delta_{12}\Delta_{13\ldots m}\Delta_{23\ldots m}=
\Delta_{3\ldots m}\Delta_{1\ldots m}$, equation (\ref{Hir3aa}) for $m$ points,
$\displaystyle{
\tau_{1\ldots m}=\frac{\det N}{\Delta_{1\ldots m}\, \tau^{m-1}}}
$,
follows from the Jacobi identity for minors of the matrix $N$:
$$
\det N \cdot \det N
\Bigl [ {\small \begin{array}{cc} 1,& \!\! m\! -\! 1\\ 2, &
\!\! m\end{array}}\Bigr ]=
\left | \begin{array}{cc} \det N [2,m] & \det N [2, m\! -\! 1]\\ &\\
\det N [1,m] & \det N [1, m\! -\! 1]\end{array}\right |.
$$

\end{document}